\documentclass[conference]{IEEEtran}

\IEEEoverridecommandlockouts
\usepackage{cite}
\usepackage{microtype}
\usepackage{amsmath,amssymb,amsfonts}
\usepackage{algorithmic}
\usepackage{graphicx}
\usepackage{tabularx}
\usepackage{textcomp}
\usepackage{xcolor}
\usepackage{threeparttable}
\usepackage{multirow}
\usepackage{booktabs}
\usepackage{float}
\usepackage{subcaption}
\usepackage{balance}

\begin{document}

\title{{An Analysis of LLM Fine-Tuning and Few-Shot Learning for Flaky Test Detection and Classification}
\thanks{This research was supported by the Natural Sciences and Engineering Research Council of Canada (NSERC), grant 2018-06588, and by MITACS.}
}


\author{\IEEEauthorblockN{Riddhi More}
\IEEEauthorblockA{\textit{Faculty of Science} \\
\textit{Ontario Tech University}\\
Oshawa, ON, Canada\\
riddhi.more1@ontariotechu.net}
\and
\IEEEauthorblockN{Jeremy S. Bradbury}
\IEEEauthorblockA{\IEEEauthorblockA{\textit{Faculty of Science} \\
\textit{Ontario Tech University}\\
Oshawa, ON, Canada\\
jeremy.bradbury@ontariotechu.ca}
}}

\maketitle

\begin{abstract}

Flaky tests exhibit non-deterministic behavior during execution and they may pass or fail without any changes to the program under test. Detecting and classifying these flaky tests is crucial for maintaining the robustness of automated test suites and ensuring the overall reliability and confidence in the testing. However, flaky test detection and classification is challenging due to the variability in test behavior, which can depend on environmental conditions and subtle code interactions. 
Large Language Models (LLMs) offer promising approaches to address this challenge, with fine-tuning and few-shot learning (FSL) emerging as viable techniques. With enough data fine-tuning a pre-trained LLM can achieve high accuracy, making it suitable for organizations with more resources. Alternatively, we introduce FlakyXbert, an FSL approach that employs a Siamese network architecture to train efficiently with limited data. To understand the performance and cost differences between these two methods, we compare fine-tuning on larger datasets with FSL in scenarios restricted by smaller datasets. Our evaluation involves two existing flaky test datasets, FlakyCat and IDoFT. Our results suggest that while fine-tuning can achieve high accuracy, FSL provides a cost-effective approach with competitive accuracy, which is especially beneficial for organizations or projects with limited historical data available for training. These findings underscore the viability of both fine-tuning and FSL in flaky test detection and classification with each suited to different organizational needs and resource availability.

\end{abstract}





\section{Introduction}

Software testing is an essential part of software development and are used to verify software works as expected and detect software bugs. The role of software testing can be potentially undermined by flaky tests that pass and fail intermittently without any changes to the program under test. Due to the fact that flaky test failures are not deterministically reproducible, developers often have to spend significant time in response to flaky test failures which ultimately may or may not be attributed back to problems in their code. 
Although it may be tempting to ignore flaky test failures, ignoring these failures can be dangerous, as they may represent real faults in the production code~\cite{Lam2019}.


Detecting and classifying flaky tests is motivated by the need to ensure software quality and development efficiency. Flaky tests can be categorized into various types based on their underlying causes, such as asynchronous waits, concurrency issues, test order dependencies and timing issues. Identifying not only the presence of flaky tests, but also the specific type of flakiness, can help developers address root causes more effectively, leading to more stable and reliable test suites~\cite{rahman2024}.

Large Language Models (LLMs) have shown significant promise in software engineering tasks such as test generation and defect prediction~\cite{WHC+24, SNET24}. Techniques such as fine-tuning and few-shot learning (FSL) enable LLMs, intended for general natural language or coding tasks, to be adapted to specific and potentially niche software engineering tasks. 
On the one hand, fine-tuning involves training a pre-trained model on a large task-specific dataset, resulting in a highly specialized model~\cite{Har24}. On the other hand, FSL allows LLMs to learn from a minimal number of examples, which makes it particularly useful for scenarios where larger task-specific datasets are not available~\cite{LVL23, Ott19, Yildirim23}. In fact, FSL is ideal for tasks such as flaky test categorization, where large annotated datasets are scarce.

FSL is especially advantageous for small organizations or development teams with limited resources and data. In such settings, the ability to achieve reasonable performance with only a few examples can be more practical and cost-effective than acquiring and labeling large datasets for fine-tuning. Additionally, FSL can be advantageous when the goal is to achieve high performance in specific and narrow tasks without the need for generalized results that can be applied in a wide range of projects~\cite{Ahmed23, Chaaben23}. This differs from fine-tuned LLMs which are typically more generalizable.

In this research, our objective is to better understand the performance and cost of fine-tuning versus FSL for flaky test detection and classification by answering the following research questions:


\begin{itemize}
    \item \textbf{RQ1: How does the performance of FSL and fine-tuning compare for flaky test detection and classification across different data scenarios?}
    \begin{itemize}
        \item \textbf{RQ1.1: What is the performance of FSL compared to fine-tuning on small per-project data?}
        The motivation behind RQ1.1 arises from 
        scenarios where the generalizability is limited and the goal is to detect or classify flaky tests in a specific project. 
        This research question seeks to explore 
        whether FSL can provide a viable alternative by effectively leveraging a small number of project-specific training examples in comparison to a generalized fine-tuned model.
        \item \textbf{RQ1.2: What is the performance of FSL compared to fine-tuning with a diverse data set?}
        RQ1.2 extends the investigation of performance to consider a comparison of FSL and fine-tuning with a varied dataset where flaky tests come from a variety of independent sources. This question is particularly pertinent in real-world applications where data heterogeneity is common. 
        The objective here is to assess how well FSL can handle high diversity in data compared to traditional fine-tuning methods. Understanding this can illuminate any potential adaptability of FSL to different contexts and the potential need for more sophisticated or tailored approaches when dealing with diverse datasets.
    \end{itemize}
    \item \textbf{RQ2:What is the cost of FSL vs. fine-tuning?}
    The performance of fine-tuning and FSL for detecting and classifying flaky tests is not done in a vacuum. It is important to view the performance differences in the context of the cost of each approach. Specifically, by comparing the training time and training data requirements of FSL and the fine-tuning approaches to the detection and classification of flaky tests, this research aims to provide insights into the overall efficiency of each method. 

\end{itemize}
By understanding the performance and cost trade-offs between FSL and fine-tuning, we aim to provide insights into the viability and appropriateness of each approach in different contexts. This evaluation and analysis can help guide practitioners in selecting the most suitable method for LLM-based flaky test detection and classification based on their specific needs and constraints.

Next we will discuss flaky tests, flaky test data sets as well as fine-tuning vs FSL methods for training LLMs (Section~\ref{sec:Background}). Then we present the existing fine-tuning and FSL approaches in the detection and classification of flaky tests (Section~\ref{sec:Related Work}). Following the existing approaches we present FlakyXbert\footnote{Source code available at: https://github.com/seer-lab/FlakyXbert}
, our new FSL-based flaky test detection and classification approach (Section~\ref{sec:FlakyXbert}). This is followed by a description of our experimental methods (Section~\ref{sec:experimental_setup_procedure}), experimental results (Section~\ref{sec:Results}) and a discussion of the trade-offs between FSL and fine-tuning in flaky test detection and classification (Section~\ref{sec:Discussion}). Lastly, we end with our conclusions and future work (Section~\ref{sec:Conclusion}).






\section{Background}
\label{sec:Background}

\subsection{Flaky Tests}

As previously mentioned, flaky tests exhibit non-deterministic behavior, often passing and failing inconsistently when run on the same code version. The consequence of this non-determinism is that it can mislead developers about the correctness of the program under test. 

Parry et al. previously investigated the consequences of flaky tests on testing reliability~\cite{Parry2021}. They found that of the 107 flaky tests identified through repeated execution of their test suites, only 50 could be reproduced as flaky when executed in isolation. This suggests that reproducing flaky tests in isolation can be challenging, although it remains useful for debugging purposes. Furthermore, their study highlights that 94 of the 96 sampled order-dependent tests caused false alarms, failing despite the absence of real bugs. 
Given these findings, 
detecting and categorizing flaky tests is crucial for several reasons. First, by identifying and isolating flaky tests, developers can focus on genuine test failures, streamlining the debugging process and reducing false alarms. Second, categorizing flaky tests can help in understanding their nature and root causes. 
Ignoring flaky test failures can significantly impact software stability~\cite{Parry2021}. 

Researchers have developed various techniques for debugging, reproducing, and repairing flaky tests, that are typically focused on specific categories of flakiness. Thus prior knowledge of the test category is crucial for effective mitigation.


Several data sets exist for studying the detection and classification of flaky tests including the International Dataset of Flaky Tests (IDoFT)~\cite{idoft2024} and the FlakyCat Dataset~\cite{Akli2023}. 

\vspace{2mm}
\begin{table}[t!]
\centering
\caption{IDoFT and FlakyCat datasets}
\label{tab:IDoFT and FlakyCat datasets}
\begin{tabularx}{\linewidth}{X r}
\midrule
{\textbf{IDoFT - Flaky vs. Non-Flaky}} & \textbf{\# Tests} \\
\midrule
Flaky tests & 3195 \\
Non-Flaky tests & 618 \\
\textbf{Total} & \textbf{3813} \\
\midrule
\multicolumn{2}{l}{\textbf{IDoFT - Flaky Test Categories}} \\
\midrule
Non-deterministic-order-dependent (NDOD)  & 84 \\
Non-order-dependent (NOD) & 226 \\
Order-dependent (OD) & 932 \\
Non-idempotent-outcome (NIO) & 196 \\
Implementation-dependent (ID) & 1617 \\
Unknown-dependency (UD) & 140 \\
\textbf{Total} & \textbf{3195} \\
\midrule
\multicolumn{2}{l}{\textbf{FlakyCat - Flaky Test Categories}} \\
\midrule
Async wait (Asyn.) & 125 \\
Concurrency (Conc.) & 48 \\
Time & 42 \\
Test Order Dependency (OD) & 103 \\
Unordered Collections (UC) & 51 \\
\textbf{Total} & \textbf{369} \\
\midrule
\end{tabularx}
\end{table}

\subsubsection{IDoFT}

The IDoFT dataset contains both flaky and non-flaky tests (see Table~\ref{tab:IDoFT and FlakyCat datasets}). This dataset was curated from a combination of smaller data sets~\cite{Shi2019,LOS+19,Wei2022} and  tests are categorized as:

\begin{itemize}
    \item \emph{order-dependent (OD):} tests whose outcomes depend on the order of test execution~\cite{LOS+19}.
    \item \emph{non-idempotent-outcome (NIO):} tests that fail when run twice in the same process and have issues with test idempotence~\cite{Wei2022}.
    \item \emph{implementation-dependent (ID):} tests that assume nondeterministic specifications (e.g., expecting unordered sets to iterate in the same order consistently)~\cite{Shi2019}.
    \item \emph{non-deterministic order-dependent (NDOD):} tests whose outcomes depends on both the order of test execution and some other non-determinism~\cite{LOS+19}.
    \item \emph{non-order-dependent (NOD):} tests whose outcomes do not depend on the order of test execution~\cite{LOS+19}.
    \item \emph{unknown dependency (UD):} tests that have non-deterministic outcomes caused by unidentified or obscure dependencies.
\end{itemize}.


\subsubsection{FlakyCat}

Luo et al. previously categorized flaky tests based on root causes including~\cite{Luo2014} :

\begin{itemize}
    \item \emph{Time:} test flakiness results in a reliance on system time (e.g., time precision failures).
    \item \emph{Concurrency (Conc):} test flakiness is due to undesirable non-deterministic thread scheduling (e.g., deadlocks, data races).
    \item \emph{Async Wait (Asyn.):} tests where flakiness relates to asynchronous calls where the result of the call is used prematurely instead of waiting for it to be available. These are technically a sub-category of concurrency.
    \item \emph{Test order dependency (OD):} the output of the test is dependent on the order of test execution.
    \item \emph{Unordered Collections (UC):} tests where assumptions are made about the order a collection's elements are returned when the elements are unordered (e.g., in sets).
\end{itemize}

This categorization was used by Akli et al.~\cite{Akli2023} in their FlakyCat dataset which classifies 369 flaky tests based on the above root causes\footnote{The FlakyCat dataset also includes a small number of flaky tests categorized on other root causes -- Network, Randomness, Test case timeout, Resource leak, Platform dependency, Too restrictive range, I/O and Floating point operations. However, these have been excluded from analysis in past research~\cite{Akli2023, rahman2024} due to the low number of data points in each category.}  (see Table~\ref{tab:IDoFT and FlakyCat datasets}).

\subsection{Large Language Models (LLMs)}

Large Language Models (LLMs) have emerged as transformative tools and revolutionizing tasks from natural language processing to code understanding. Based on the Transformer architecture introduced by Vaswani et al.~\cite{Vaswani2017Attention}, LLMs including BERT~\cite{Jacob2018BERT} and CodeBERT~\cite{Guo2020CodeBERT} have significantly enhanced automation in software development.

BERT is general purpose and focused on contextual text understanding while CodeBERT is a pre-trained version of BERT intended for code interpretation and code generation tasks. LLMs including CodeBERT have been used to improve automated code review, bug fixing~\cite{Zhou2019}, and test case generation~\cite{Siddiq2024LLMSoftware}. In the context of flaky test detection and classification, CodeBERT's dual proficiency in code and natural language enables analysis of test code structures and potential flakiness patterns.

Beyond task automation, LLMs can contribute to software reliability and efficiency. For example, Pan et al. demonstrate CodeBERT's use in defect prediction~\cite{pan2021CodeBERTSurvey} while Wang et al. overview LLM applications in software testing~\cite{WHC+24} and Schäfer et al. show LLMs' efficiency in unit test generation~\cite{Schäfer2024}.

While LLMs have been demonstrated as beneficial in Software Engineering, their deployment raises concerns about computational resource requirements and sustainability. Luccioni et al. highlight these issues by estimating the carbon footprint of training LLMs and encouraging more sustainable practices in artificial intelligence (AI) development~\cite{LVL23}.

As we explore LLMs for flaky test detection and classification, we must balance their technical capabilities with implementation constraints in real-world software development. Our research is motivated by the need for an improved understanding of the trade-offs between fine-tuning and few-shot learning (FSL) techniques in addressing flaky test challenges. Our work aims to contribute to the discourse on responsible and efficient AI utilization in Software Engineering, particularly in enhancing test suite reliability and maintainability.

\subsubsection{Fine-tuning}
Within Software Engineering, fine-tuning is an essential technique in the application of Large Language Models (LLMs) like BERT~\cite{Jacob2018BERT} and GPT~\cite{OpenAI2023}. This is particularly true for tasks such as test generation, defect prediction, and flaky test categorization~\cite{WHC+24, SNET24}. Fine-tuning involves adapting a pre-trained model, which has initially learned from a broad dataset, to a narrower, task-specific dataset that encapsulates the particularities of software development challenges. For instance, in flaky test categorization, fine-tuning helps models to accurately distinguish between stable and unstable tests by training on labeled datasets that define tests as flaky or non-flaky, thus addressing the critical issue of test reliability in automated testing environments~\cite{Ott19, Yildirim23}.

The utility of fine-tuning extends beyond just identifying flaky tests, is observed with other software engineering tasks including bug prediction and automatic code reviews~\cite{Bromley1993, Akli2023}. Through fine-tuning, LLMs leverage their substantial pre-trained knowledge base, refining it against specific datasets to align closely with the complex requirements of software systems. Fine-tuning not only enhances model precision but also ensures that the models are robust~\cite{Jacob2018BERT, WHC+24}. 

\subsubsection{Few-Shot Learning}
Few-shot learning (FSL) has evolved significantly since its early implementations in fields like image recognition and has increasingly found relevance in software engineering, particularly in environments constrained by data availability or computational resources~\cite{Bromley1993}. Unlike fine-tuning, which relies on large volumes of task-specific data, FSL can achieve competitive results with significantly fewer data points by exploiting the pre-trained capabilities of models like BERT and GPT. This quality makes FSL exceptionally advantageous for startup environments or in scenarios where rapid deployment is crucial, without the lengthy and resource-intensive retraining phases typical of fine-tuning approaches~\cite{Jacob2018BERT, OpenAI2023}.

Historically, FSL's origins can be traced back to cognitive science efforts to mimic human learning efficiency, where humans often learn from only a few examples. This paradigm was initially utilized with techniques like the support vector machines and later gained traction with more complex architectures such as Siamese networks and transformer-based models~\cite{Bromley1993, Jacob2018BERT}. In software engineering, FSL has been deployed in anomaly detection, where models learn to identify issues from a small set of examples of anomalous code, and in automated code generation, where the system suggests code snippets based on a limited context~\cite{Jacob2018BERT,WHC+24}.

The strength of FSL lies in its flexibility and the lower data needs, which contrasts sharply with the traditional fine-tuning methods that often require extensive labeled datasets that are not only expensive to create but also difficult to obtain in many real-world scenarios. Moreover, FSL aligns well with the ongoing shift towards more agile and iterative development practices in software engineering, supporting rapid adaptation to new tasks without the need for extensive retraining. This has not only opened up new avenues for deploying advanced AI in smaller-scale projects but has also significantly reduced the time and cost associated with developing intelligent software solutions~\cite{Ott19,Yildirim23}.

\section{Related Work}
\label{sec:Related Work}

Detecting and classifying flaky tests are critical for maintaining the reliability of software test suites. Various techniques have been developed to address these tasks, each with its strengths and limitations.
\vspace{2mm}
\begin{figure*}[t!]
    \centering
    \includegraphics[width=0.9\textwidth]{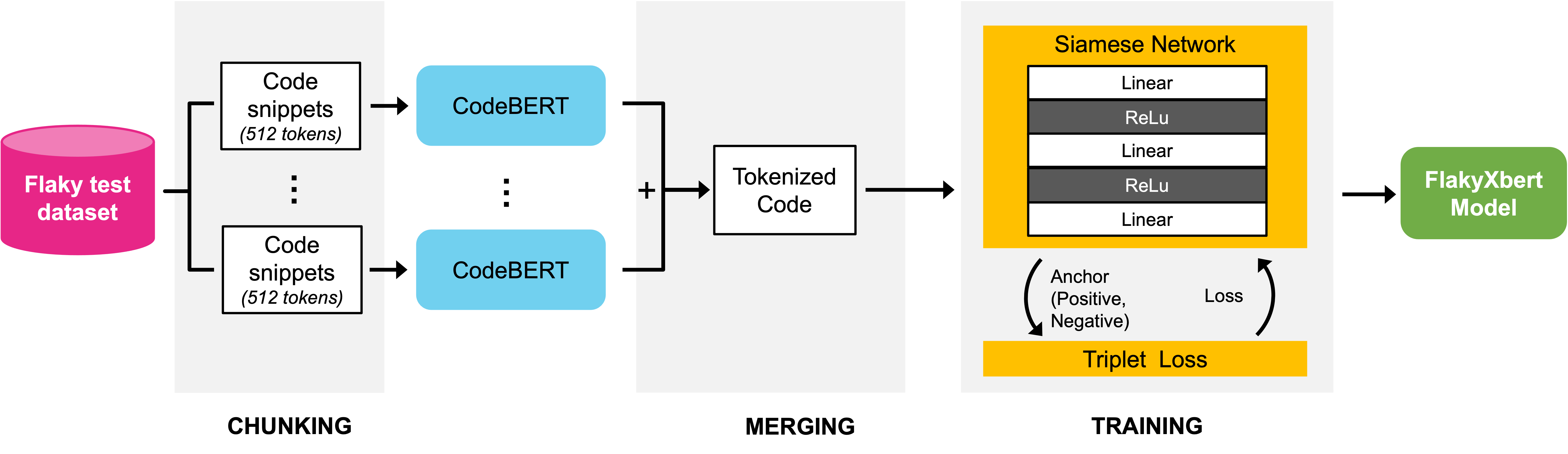} 
    \caption{Architecture of FlakyXbert}
    \label{fig:flakyxbert}
\end{figure*}

\subsection{Detection Techniques}

One of the prominent tools for flaky test detection is FlakeFlagger~\cite{Alshammari21}. FlakeFlagger employs machine learning models to predict flaky tests by analyzing features extracted from test execution logs and results. This approach has achieved promising results across multiple datasets by identifying patterns and anomalies indicative of flakiness. However, FlakeFlagger relies on extensive feature engineering and access to detailed execution logs, which may not always be available.

Another innovative approach is Flakify, a black-box, language model-based predictor for flaky tests that relies exclusively on the source code of test cases. Flakify utilizes CodeBERT~\cite{Guo2020CodeBERT}, a pre-trained language model, which is fine-tuned to predict flaky tests based on code features. This eliminates the need for access to production code and rerunning test cases, making it a practical solution for various settings. However, the dependency on source code alone might limit its accuracy in certain contexts~\cite{fatima2023}.

Shanto Rahman et al.,~\cite{rahman2024} have introduced a method for quantizing large language models to predict flaky tests. Quantization aims to optimize model size and performance, allowing it to operate with reduced computational resources while maintaining high accuracy. This approach is particularly beneficial for reducing computational costs during prediction. To address potential accuracy losses due to quantization, Rahman et al. integrated additional classifiers, such as random forests, to enhance performance post-quantization.

\subsection{Classification Techniques}

Classifying flaky tests into specific categories is crucial for understanding their root causes and effectively addressing them. FlakyCat is a technique that categorizes flaky tests based on their root causes, including asynchronous waits, concurrency issues, and test order dependencies. By using few-shot learning, FlakyCat can classify tests with minimal labeled data, making it practical for various projects. This approach allows developers to pinpoint the specific type of flakiness and address it accordingly~\cite{Akli2023}.

In addition to detection, Rahman’s work extends to the classification of flaky tests. By leveraging large language models, their approach aims to categorize flaky tests into meaningful groups to facilitate debugging and resolution. This classification helps in understanding the patterns and causes of flakiness across different projects, providing valuable insights for improving test reliability~\cite{rahman2024}.

\section{Architecture of FlakyXbert}
\label{sec:FlakyXbert}

Building upon the pioneering work of FlakyCat, we introduce FlakyXbert, a novel architecture that incorporates few-shot learning techniques to enhance the detection of flaky tests. FlakyXbert refines and extends the computational methodologies originally developed in FlakyCat, adapting these to harness the power of few-shot learning for more precise and efficient test analysis. This section provides a detailed outline of the FlakyXbert architecture, ensuring clarity and reproducibility for future research and implementation efforts. Fig.~\ref{fig:flakyxbert} illustrates the comprehensive architecture of FlakyXbert, serving as a visual guide to the enhanced design and functionality of the model.

\subsection{Data Preparation and Tokenization}
The initial stage of the FlakyXbert pipeline involves meticulous data preparation, essential for ensuring the quality and consistency of input data for model training. Each test case is first segmented into chunks. These chunks are then tokenized using the AutoTokenizer component from the `Microsoft/codebert-base` model, which is designed specifically for handling programming languages.

After tokenization, the tokenized segments are reassembled back to their original sequence to maintain the structural and contextual integrity of the test cases. This step is crucial for preserving the flow and meaning of the code within each test case. Once reassembled, padding is applied after tokenization to ensure all token sequences have the same length. Shorter sequences are padded with zeroes up to the length of the longest sequence in the batch. The padding process ensures that all sequences are extended to the length of the longest sequence in the dataset. This uniformity is vital for the model to process the input data efficiently.

This approach, while introducing some overhead, is significantly more efficient than fine-tuning processes that require similar steps. By focusing on efficient data preparation, FlakyXbert minimizes computational expenses while maintaining high data quality, which is instrumental for the effective learning and performance of the model.

\subsection{Siamese Network Design}
At the heart of FlakyXbert lies a sophisticated Siamese neural network, enhanced from the FlakyCat architecture to include several key innovations:

\begin{itemize}
\item The network utilizes twin branches, each comprising a sequence of convolutional and fully connected layers. These branches process paired inputs — consisting of an anchor, a positive example (similar to the anchor), and a negative example (dissimilar to the anchor) — to generate embeddings that effectively capture the characteristics of flaky versus non-flaky tests.

\item The embeddings are optimized using a triplet loss function, which plays a crucial role in the learning dynamics of the model. Mathematically this function is defined as:
    \begin{equation*}
        \text{TL} = \max(\|f(a) - f(p)\|^2 - \|f(a) - f(n)\|^2 + \text{margin}, 0)
    \end{equation*}
    Here, \(f(x)\) represents the embedding of input \(x\). The terms \(a\), \(p\), and \(n\) denote the anchor, positive, and negative inputs, respectively. In this context, the positive input (\(p\)) is an example that is similar to the anchor (\(a\)), and the negative input (\(n\)) is dissimilar. The function seeks to minimize the distance between the anchor and the positive example (making them closer in the embedding space) while maximizing the distance between the anchor and the negative example (pushing them further apart). This process is facilitated by the inclusion of a 'margin', a threshold value that quantifies the minimum desired difference between the positive and negative distances. The margin is finely tuned to ensure optimal separation between classes in the embedding space, effectively enhancing the model’s ability to discriminate between different categories. Positive and negative examples are selected randomly from the dataset, providing a diverse range of comparisons to robustly train the model.
\end{itemize}

\vspace{2mm}
\begin{figure*}[htbp]
    \centering

     \begin{subfigure}[b]{17cm}
         \centering
         \includegraphics[width=\textwidth]{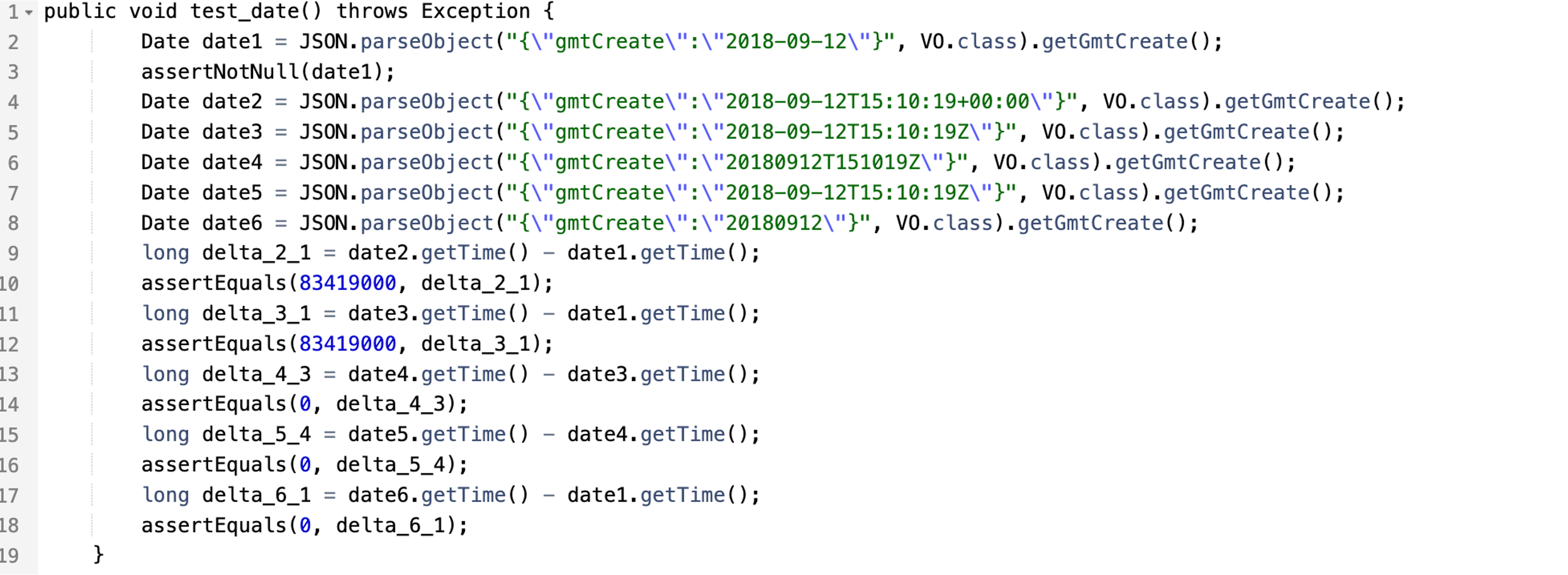}
         \caption{ testing date parsing and difference calculation between parsed dates}
         \label{timing1}
     \end{subfigure}
      \centering
 \begin{subfigure}[b]{17cm}
\vspace{2mm}
         \centering
         \includegraphics[width=\textwidth]{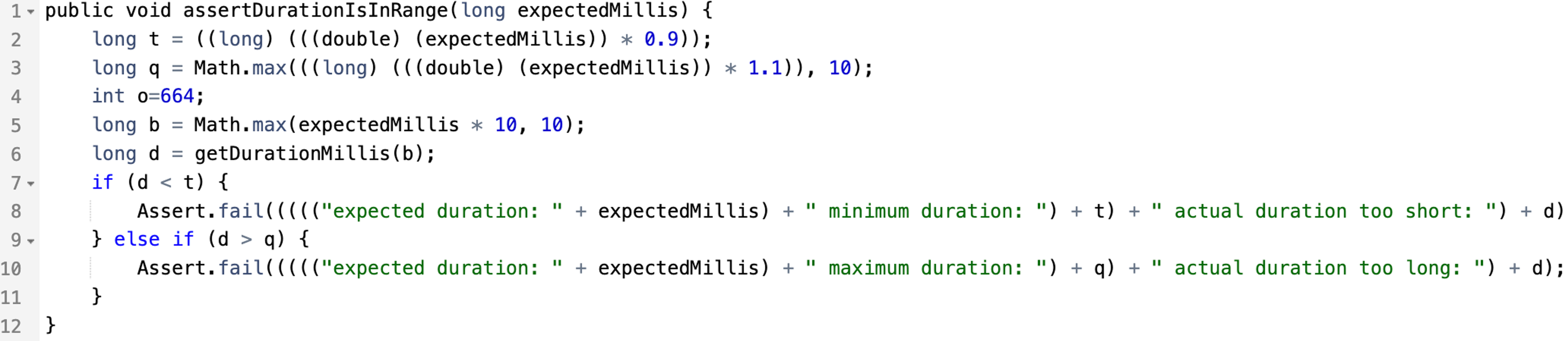}
         \caption{ testing a computed duration falls within an expected range}
         \label{timing2}
     \end{subfigure}
    
    \caption{Example of High variation in FlakyCat test data for tests classified as 'Time'}
\vspace{1mm}
    \footnotesize{\emph{Comparative Analysis of Time-Related Testing in Software. The top snippet demonstrates testing date parsing and difference calculation between parsed dates, while the bottom snippet verifies that a computed duration falls within an expected range, illustrating the varied testing methodologies within the same 'time' category and the inherent challenges in model training due to these differences.}}
    \label{fig:time_testing}
\end{figure*}

\subsection{Optimization and Training}
FlakyXbert is trained using stochastic gradient descent with a backpropagation algorithm tailored for triplet loss optimization. The network is trained iteratively, wherein each epoch, the model parameters are updated to minimize the triplet loss across all training examples. The learning rate and other hyperparameters (see Section~\ref{setup}) are carefully selected based on preliminary experiments to ensure convergence and optimal performance.

\subsection{Implementation Details}
The implementation leverages PyTorch for model development, utilizing its robust ecosystem for deep learning research. The training process is expedited by employing GPU acceleration, allowing for the processing of large datasets and complex model computations efficiently. For the tokenization and initial embedding of code snippets, FlakyXbert integrates the Hugging Face Transformers library, utilizing the pre-trained CodeBERT model to ensure state-of-the-art performance.

\subsection{Validation and Performance Evaluation}
Model validation is performed on a held-out test set, split from the original dataset before training. The model's performance is assessed based on its ability to correctly classify new, unseen test cases. Metrics such as precision, recall, F1 score, and a confusion matrix are computed to provide a comprehensive evaluation of the model's effectiveness.

\begin{table*}[ht]
 \centering
  \small
{\renewcommand{\arraystretch}{1.2}
\caption{Comparison between FlakyXbert and FlakyCat}
\label{tab:comparison}
\begin{tabular}{@{}l|l|l@{}}
\toprule
\textbf{Characteristic} & \textbf{FlakyXbert} & \textbf{FlakyCat} \\ \midrule
Normalization & No explicit normalization of outputs & L2 normalization of outputs \\
Architecture & Siamese network (with linear layers, ReLU activations) & Single dense layer with 512 units \\
Output Dimensionality & Output dimensionality equals \textit{embedding\_size} & 512-dimensional output \\
Network Capacity & High (with multiple layers and nonlinear activations) & Low (due to single layer) \\
Non-linearity & ReLU non-linearities between layers & No non-linear activations  \\
Feature Compression & Potential for more feature compression due to bottleneck & Less feature compression \\
Expected Output Variety & High (due to non-linear transformations and higher capacity) & Low (due to linear transformation) \\ \bottomrule
\end{tabular}
}
\end{table*}

\subsection{FlakyXbert vs. FlakyCat}



FlakyCat~\cite{Akli2023}, previously discussed in Section~\ref{sec:Related Work}, has made a novel contribution to the field of flaky test detection, utilizing an overlapping sliding window technique for data segmentation and a single dense layer architecture. Building on this foundation, FlakyXbert introduces key advancements, summarized in Table~\ref{tab:comparison}, such as a Siamese network with triplet loss and non-linear transformations using ReLU activations. 
FlakyXbert’s architecture allows for more sophisticated pattern learning and greater adaptability in few-shot learning scenarios, making it particularly effective in data-scarce environments.

\vspace{5mm}
\begin{table}[t]
\centering
\caption{F1-score Comparison on Flaky vs Non-Flaky Test Detection on IDoFT dataset}
\label{tab:idoftFvNF}
\resizebox{\columnwidth}{!}{\begin{tabular}{l|r|r|r|r|r}
\toprule
Project & Support & FlakyXbert & Flakify++ & Q-Flakify++ & FlakyQ\_RF \\
\midrule
Dubbo & 186 & 88.7 & 87.0 & 91.0 & 88.0 \\
Hadoop & 149 & 95.0 & 99.0 & 100.0 & 100.0 \\
Nifi & 146 & 91.5 & 99.0 & 100.0 & 100.0 \\
Junit & 250 & 94.0 & 99.0 & 99.0 & 99.0 \\
Admiral & 113 & 91.3 & 99.0 & 99.0 & 99.0 \\
Fastjson & 109 & 91.3 & 91.0 & 93.0 & 93.0 \\
spring & 68 & 100.0 & 100.0 & 100.0 & 100.0 \\
Adyen & 89 & 30.0 & 43.0 & 52.0 & 45.0 \\
Mockserver & 39 & 100.0 & 100.0 & 100.0 & 100.0 \\
\midrule
Total/\\Weighted Avg. & 2105 & 95.1 & 95.6 & 96.0 & 95.4 \\
\bottomrule
\end{tabular}}
\begin{tablenotes}
\item \footnotesize{\emph{Note: To see the full version, please refer to the original paper~\cite{rahman2024}.}}
\end{tablenotes}
\end{table}

\begin{figure}[h] 
    \centering
    \includegraphics[width=0.5\textwidth]{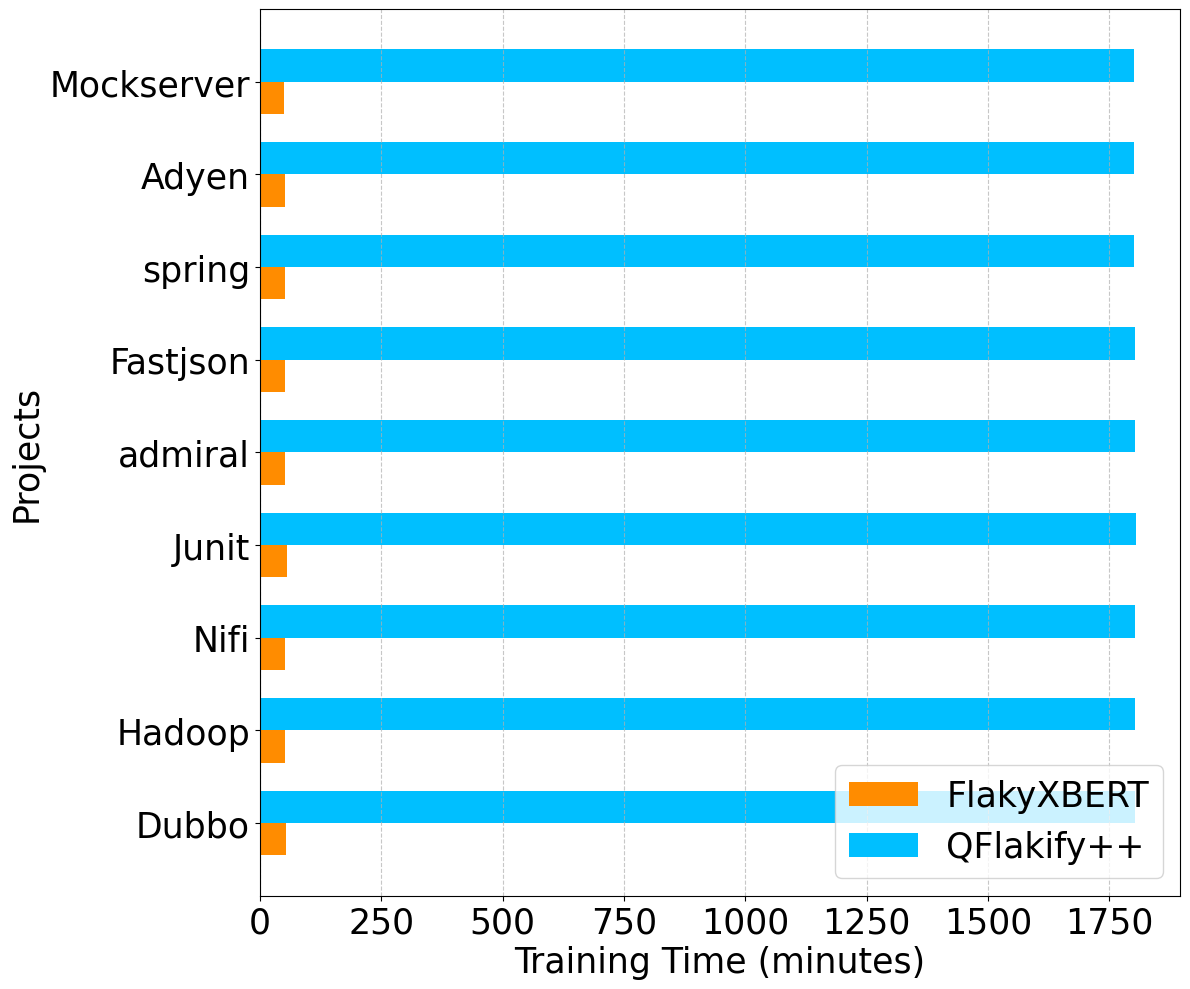} 
    \caption{Training Time comparison: FlakyXbert vs QFlakify++ on IDoFT per-project detection of Flaky vs Non-Flaky Tests}
    \vspace{1mm}
    \footnotesize{\emph{}}
    \label{fig:costsidoftdetect}
\end{figure}

\section{Experimental Setup \& Procedure}
\label{sec:experimental_setup_procedure}

\subsection{Data}

Our study utilizes two primary datasets: the International Dataset of Flaky Tests (IDoFT) and the FlakyCat Dataset. The use of these datasets is directly tied to our research questions:

\begin{itemize}
    \item \textbf{IDoFT Dataset}: This dataset contains both flaky and non-flaky tests from various projects. We use it to address RQ1.1, which focuses on per-project performance on smaller datasets\footnote{For IDoFT, we applied an inclusion criterion that only projects with at least three examples per flaky test category were considered. This criterion ensures a minimum level of diversity within each category.}

    \item \textbf{FlakyCat Dataset}: This dataset provides a diverse collection of flaky tests categorized by their root causes. We use it to address RQ1.2, which examines performance on highly diverse data. 
\end{itemize}

Both datasets contribute to answering RQ2 by providing different scales and complexities of data, which impacts the computational resources required for training.

\subsection{Metrics}
To comprehensively evaluate the performance and cost of FSL versus fine-tuning, we employ the following metrics\footnote{While precision and recall metrics were also analyzed and yielded trends consistent with the F1 score, only the F1 score results are presented here.}:

\begin{itemize}
        \item \textbf{F1-score:} The harmonic mean of precision and recall, providing a balanced measure of the model's accuracy.
        This metrics allow us to assess the effectiveness of both FSL and fine-tuning in detecting and classifying flaky tests, directly addressing RQ1.1 and RQ1.2.

\item \textbf{Computation Time}: The computation time required to train each model is used as the primary cost metric in this study. This measures the total time taken for the models to complete their training processes, providing insight into the computational efficiency of each approach.
This allows us to answer RQ2.

\end{itemize}


\subsection{Experimental Setup:}
\label{setup}
The experiments on both the FlakyCat and IDoFT dataset were conducted using the FlakyXbert model, which integrates a Siamese network architecture tailored for few-shot learning, this model was specifically designed to handle the sparse representation of flakiness categories in datasets where examples of certain types of flakiness are limited. The FlakyXbert was trained for 450 epochs using a contrastive loss function, which effectively distinguishes between the different root causes of flakiness by comparing pairs of test cases.

For training, the learning rate was set to \(1 \times 10^{-5}\) and the batch size was 8 to optimize the balance between training speed and memory usage. These hyper-parameters were chosen to enhance the model's ability to learn nuanced features from a small number of training examples without over-fitting.

In the IDoFT dataset, each project was treated as a separate entity, allowing for customization of the data handling and model training to match the unique testing environments of each project. A stratified test-train split was employed for each project to ensure that all types of flakiness present were proportionally represented in both training and test sets. We focused on per-project evaluation for the IDoFT dataset because few-shot learning (FSL) is designed for scenarios with small, localized datasets, such as those within a single project. 

The Flakify++, Q-Flakify++ and FlakyQ\_RF models 
all perform the per-project evaluation differently, reserving the data from one project for use as the test set while using the data from all the other projects for training. 
The decision to use a different training setup for the fine-tuned models versus the FSL model was a deliberate choice as we wanted to compare each model based on its intended use with respect to generalizability and training data needs. 






    
\vspace{5mm}
\begin{table}[t]
\centering
\caption{F1-Score Comparison on Flaky Test Classification on IDoFT Dataset}
\label{tab:idoftFlakycategory}
\begin{threeparttable}
\resizebox{\columnwidth}{!}{\begin{tabular}{l|r|r|r|r|r}
\toprule
Project & Support & FlakyXbert & Flakify++ & Q-Flakify++ & FlakyQ\_RF \\

\midrule
Dubbo & 170 & 71.0 &77.0 & 77.0 &  73.0 \\
Hadoop & 146 & 51.0 & 90.0 & 88.0 & 91.0 \\
Nifi & 139 & 91.0 & 100.0 & 100.0 & 100.0 \\
Junit & 250 & 94.0 &98.0 & 98.0 & 98.0 \\
Ormlite & 113 & 96.0 & 99.0 & 97.0 & 97.0 \\
admiral & 109 & 63.0 & 85.0 & 77.0 & 88.0 \\
Wildfly & 84 & 74.0 & 97.0 & 98.0 & 98.0 \\
Mapper & 75 & 100.0 & 93.0 & 80.0 & 100.0 \\
Fastjson & 64 & 82.0 & 91.0 & 88.0 & 94.0 \\
Java & 54 & 85.0 & 87.0 & 87.0 & 87.0 \\
Biojava & 51 & 91.0 & 19.0 & 16.0 & 32.0 \\
spring  & 68 & 90.0 & 100.0 & 100.0 & 100.0 \\
Hbase  & 47 & 76.0 & 98.0 & 95.0 & 98.0 \\
hive & 41 & 100.0 & 98.0 & 96.0 & 98.0 \\
Nacos & 32 & 96.0 & 100.0 & 97.0 & 97.0 \\
\midrule
Total/\\Weighted Avg. & 1810 & 76.5 & 90.2 & 93.0 & 94.8 \\
\bottomrule
\end{tabular}}
\begin{tablenotes}
\item \footnotesize{\emph{Note: To see the full version, please refer to the original paper~\cite{rahman2024}}}
\end{tablenotes}
\end{threeparttable}
\end{table}

\begin{figure}[t] 
    \centering
\includegraphics[width=0.49\textwidth]{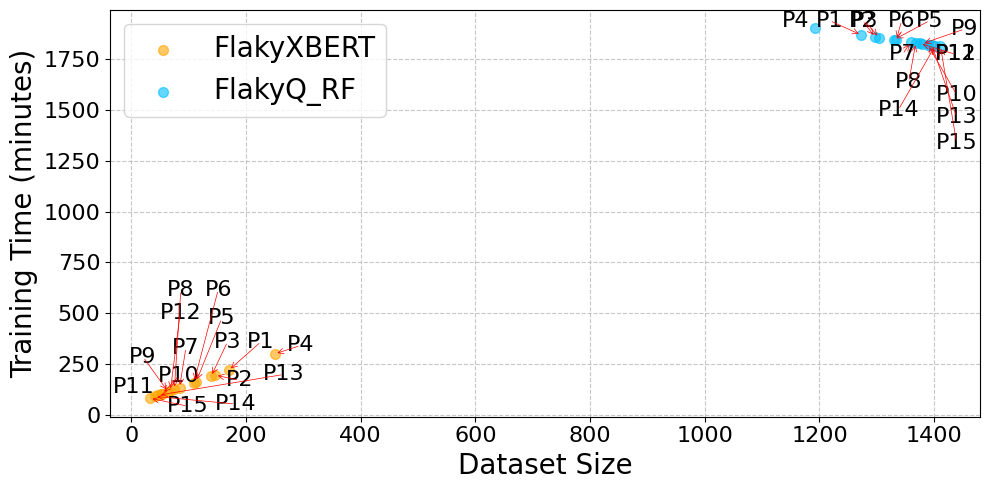} 
    \caption{Training time and dataset size: FlakyXbert vs FlakyQ\_RF on IDoFT per-project classification of flaky test categories}
    \vspace{1mm}
    \footnotesize{\emph{}}
    \label{fig:costsidoftclass}
\end{figure}

\subsection{Hardware Configuration}

The experiments were conducted on a Dell RTX 4090 workstation with Nvidia drivers (ver. 555) and CUDA~12.5. 

\section{Results}
\label{sec:Results}

We evaluate FlakyXbert's performance in detecting and classifying flaky tests using the IDoFT and FlakyCat datasets that answer RQ1 and RQ2. Our analysis focuses on project-wise performance for the IDoFT dataset and overall performance for the FlakyCat dataset. The results for the IDoFT dataset are combined into Table~\ref{tab:idoftFvNF} \&~\ref{tab:idoftFlakycategory} for detection and classification respectively where each row corresponds to a different classifier. The columns “P”, “R”, and “F1” show precision, recall, and F1-score, respectively.  A comparison study is done with FlakyCat proposed by Akli et al., and Flakify++, QFlakify++, and other variants called the QFlakify classifiers~\cite{rahman2024} that leverage K-nearest neighbour, Random forest, SVM, and more proposed by  Rahman et al.,. The table titled “Performance Comparison on Flaky Test Categorization on FlakyCat Dataset" Table~\ref{tab:flakycatCategory} showcases the effectiveness of various classifiers under three distinct techniques: Few-shot Learning (FSL), Fine-tuning (FT), and a Hybrid of FSL and FT across these same models.

\begin{figure}[ht] 
    \centering
    \includegraphics[width=0.49\textwidth]{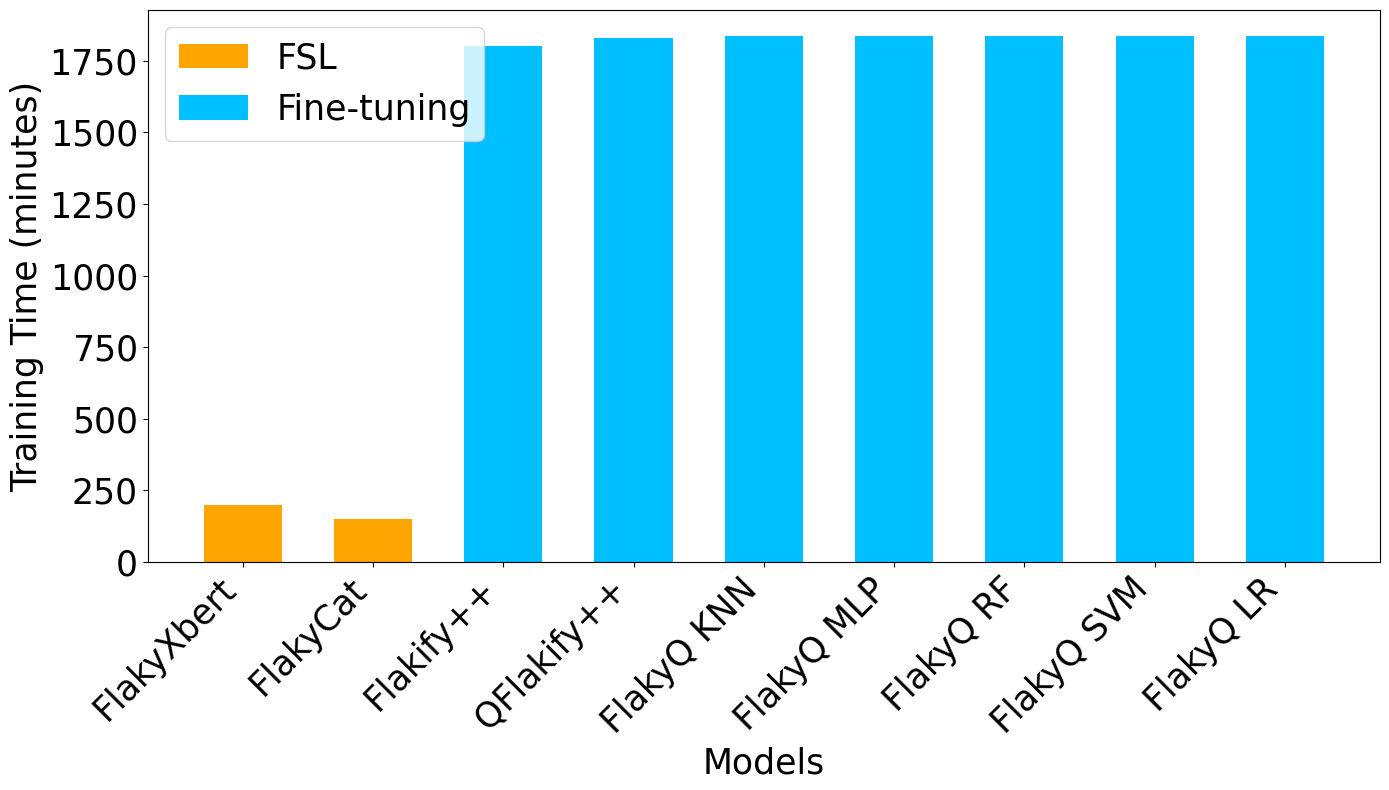} 
    \caption{Training time comparison: FSL vs Fine-tuning on FlakyCat Dataset for classification of flaky test categories}
    \vspace{1mm}
    \footnotesize{\emph{}}
    \label{fig:costsflakycat}
\end{figure}

The Fig.~\ref{fig:costsidoftdetect},~\ref{fig:costsidoftclass}~\&~\ref{fig:costsflakycat} presented provides empirical evidence addressing RQ2, focusing on the cost comparison in terms of training time between Few-Shot Learning (FSL) and traditional fine-tuning methodologies. By illustrating the training time differences across various models derived from a baseline, we assess the computational expense incurred when employing FSL as opposed to extensive fine-tuning.

\subsubsection{\textbf{Detection of Flaky Tests}}
We first assessed FlakyXbert's capability to detect flaky tests on a per-project basis using the IDoFT dataset.

The graph in Fig.~\ref{fig:costsidoftdetect} presents a comparison of training times between the FlakyXbert model and the QFlakify++ model across several projects from the IDoFT dataset as they are the best-performing ones from the FSL and Fine-tuning category respectively. The graph clearly shows a substantial difference in training duration between the two models, with FlakyXbert consistently requiring significantly less time to complete training than  QFlakify++. Across all projects, including Mockserver, Adyen, Spring, Fastjson, Admiral, Junit, Nifi, Hadoop, and Dubbo, the training time for  QFlakify++ exceeds 1700 minutes, while FlakyXbert completes the training process within a fraction of this time, often under 200 minutes.

Despite the differences in training time, Table~\ref{tab:idoftFvNF} shows that FlakyXbert maintains strong performance in terms of precision, recall, and F1-scores. For example, both FlakyXbert and QFlakify++ achieve perfect F1-scores on projects like Mockserver, Nifi, and Junit. However, FlakyXbert shows a notable drop in performance for Adyen (F1 of 30.00 compared to 45.00 for ), though the training time difference remains a key advantage. This balance between computational efficiency and solid performance makes FlakyXbert an appealing option for faster model deployment.
\vspace{2mm}
\begin{table*}[ht]
\centering
\caption{F1-Score Comparison on Flaky Test Categorization on FlakyCat Dataset}
\label{tab:flakycatCategory}
\begin{threeparttable}
\begin{tabular}{l|l|rrrrr|r}
\toprule
\textbf{Technique} & \textbf{Classifier} & \textbf{Asyn.} & \textbf{Conc.} & \textbf{Time} & \textbf{UC} & \textbf{OD} & \textbf{Weighted Avg.} \\
\midrule
\multirow{2}{*}{Few-shot Learning (FSL)
} & FlakyXbert & 98.0 & 90.0 & 93.0 & 97.0 & 99.0 & 96.0 \\
& FlakyXbert (without augmentation) & 52.0 & 80.0 & 36.0 & 43.0 & 78.0 & 60.0 \\
 & FlakyCat & 72.0 & 36.0 & 75.0 & 72.0 & 73.0 & 67.5 \\
\midrule
\multirow{7}{*}{Fine-tuning (FT)} & Flakify++ & 94.8 & 93.3 & 96.9 & 96.1 & 97.1 & 95.6 \\
 & Q-Flakify++ & 92.6 & 87.1 & 96.9 & 95.0 & 95.8 & 93.6 \\
 & FlakyQ\_KNN & 93.1 & 90.7 & 95.5 & 95.0 & 96.3 & 94.2 \\
 & FlakyQ\_MLP & 94.0 & 89.7 & 95.5 & 94.8 & 96.6 & 94.5 \\
 & FlakyQ\_RF & 94.3 & 91.5 & 95.5 & 94.8 & 96.6 & 94.8 \\
 & FlakyQ\_SVM & 93.8 & 89.0 & 95.5 & 93.2 & 96.1 & 93.9 \\
 & FlakyQ\_LR & 92.7 & 89.8 & 95.5 & 94.8 & 96.1 & 93.9 \\
\midrule
Hybrid (FSL + FT) & FSL++ & 93.7 & 90.3 & 97.9 & 95.9 & 96.7 & 91.5 \\
\bottomrule
\end{tabular}
\begin{tablenotes}
\item \footnotesize{\emph{Note: Asyn. = Async Wait, Conc. = Concurrency, Time = Test Order Dependency, UC = Unordered Collections, OD = Other Dependencies}}
\end{tablenotes}
\end{threeparttable}
\end{table*}

\vspace{2mm}

\subsubsection{\textbf{Classification of Flaky Tests}}
We further evaluated FlakyXbert's ability to classify flaky tests into specific categories using both the IDoFT and FlakyCat datasets.

\paragraph{\textbf{IDoFT Dataset Classification}}
Fig.~\ref{fig:costsidoftclass} compares the training time against the dataset size for both FlakyXbert and FlakyQ\_RF across multiple projects in the IDoFT dataset. Similar to the detection task, FlakyXbert demonstrates a clear advantage in terms of training efficiency, requiring substantially less time to train across varying dataset sizes. For smaller projects, such as P1 and P9, FlakyXbert completes training well under 250 minutes, while FlakyQ\_RF consistently demands much longer durations -- exceeding 1000 minutes in cases where dataset sizes grow, such as in P12 and P13. This highlights FlakyXbert’s scalability, especially for smaller projects, where it offers improved training times without sacrificing performance.

Table~\ref{tab:idoftFlakycategory} presents the classification performance of FlakyXbert, FlakyQ\_RF, and other models (Flakify++ and Q-Flakify++). FlakyXbert maintains strong precision, recall, and F1-scores across various projects, achieving perfect or near-perfect results on projects such as Nifi (F1 = 91.0), Mapper (F1 = 100.0), and Ormlite (F1 = 96.0). While FlakyQ\_RF outperforms FlakyXbert in some projects, such as Hadoop (F1 = 91.0 vs. 51.0 for FlakyXbert), it consistently requires significantly longer training times, as shown in Fig.~\ref{fig:costsidoftclass}.

The balance between training time and classification performance is evident when comparing the weighted averages in Table~\ref{tab:idoftFlakycategory}. FlakyQ\_RF achieves a higher overall F1-score (94.8) compared to FlakyXbert (76.5), but this comes at the cost of greatly increased computational time, as highlighted in Fig.~\ref{fig:costsidoftclass}. For scenarios where training efficiency is paramount, particularly in smaller or time-sensitive projects, FlakyXbert proves to be a highly competitive alternative to FlakyQ\_RF.

\paragraph{\textbf{FlakyCat Dataset Classification}}

The initial performance of FlakyXbert on the FlakyCat dataset, composed of diverse test cases without a discernible pattern, was markedly inconsistent. The variability is evident in the substantial differences in performance metrics, as shown in Table~\ref{tab:flakycatCategory}. This inconsistency can be attributed to the dataset's heterogeneity, which poses a challenge for few-shot learning (FSL) models like FlakyXbert (see Fig.~\ref{fig:time_testing}). 

To mitigate this challenge and improve FlakyXbert's generalization across diverse data, a data augmentation strategy inspired by the Synthetic Minority Over-sampling Technique (SMOTE)~\cite{Chawla2002} was employed. This technique involved duplicating existing test cases and mutating non-critical elements—such as variable names, constants, and data types—using the Spoon library, as described by Akli et al.~\cite{Akli2023}. These mutations created 639 unique test scenarios while preserving the test behavior, enhancing the model's training robustness by providing a more diverse yet structurally similar dataset.

Without augmentation, FlakyXbert performed poorly, achieving only 52\% in Asynchronous Wait, 80\% in Concurrency, 36\% in Time, 43\% in Unordered Collections, and 78\% in Other Dependencies, with a weighted average of just 60\%. After applying data augmentation, the model's performance improved significantly across all categories, achieving 98\% in Asynchronous Wait, 90\% in Concurrency, 93\% in Time, 97\% in Unordered Collections, and 99\% in Other Dependencies, with a weighted average of 96\%. This improvement, evident in Table~\ref{tab:flakycatCategory}, highlights the effectiveness of FSL techniques when combined with data augmentation strategies like SMOTE.

The results demonstrate that enriching the dataset through synthetic data generation broadens the learning scope of FlakyXbert, enabling the model to better capture the nuanced characteristics of flaky tests. This combination of FSL with augmentation strategies allows the model to handle the inherent diversity of the FlakyCat dataset with greater precision and accuracy.

Fig.~\ref{fig:costsflakycat} shows the cost comparison in terms of training time. Although FSL models like FlakyXbert and FlakyCat provide a clear advantage in terms of faster training (under 250 minutes), the original FlakyCat model~\cite{Akli2023} suffers from a substantial drop in F1-score by 28.5\%, with only a marginal reduction in training time (13 minutes). Fine-tuning models such as Flakify++ and Q-Flakify++ show smaller declines in F1 scores (-13\% and -0.4\%, respectively) while requiring significantly more training time (1667 and 1672 minutes). Other models like FlakyQ KNN, MLP, RF, SVM, and LR exhibit slight reductions in F1 scores (-1.2\% to -2.1\%) but also incur high computational costs.

In conclusion, while fine-tuning models consistently achieve high performance, they do so at the expense of substantial training time. In contrast, FlakyXbert demonstrates that the combination of FSL and data augmentation strategies like SMOTE can drastically reduce training time while maintaining competitive accuracy, making it a highly efficient alternative for flaky test categorization.







\section{Discussion}
\label{sec:Discussion}

Our empirical evaluation reveals significant insights into the effectiveness and cost-efficiency of few-shot learning (FSL) and fine-tuning methodologies in the context of flaky test detection and classification. Recall that our research questions are: 

\textbf{RQ1: How does the performance of FSL and fine-tuning compare for flaky test detection and classification across different data scenarios?}

\textbf{RQ1.1: What is the performance of FSL compared
to fine-tuning on small per-project data?}

Our research evaluates the efficacy of FSL against  fine-tuning techniques within the context of small datasets. The evidence demonstrates that FSL not only matches but occasionally surpasses fine-tuning in accuracy on a per-project basis. FSL's proficiency in learning from a reduced number of examples provides a distinct advantage in situations where access to training data is limited or the cost of training data acquisition is high.

\textbf{RQ1.2: What is the performance of FSL compared to fine-tuning with a diverse data set?}

Our investigation into RQ1.2 reveals that while FSL generally maintains robust performance across diverse datasets, we observed some inconsistencies in performance under certain conditions. To address this variability and enhance the reliability of FSL, we implemented code augmentation techniques. These methods involve generating augmented data that mimics the diversity of the original dataset but with reduced variability, ensuring more uniform and predictable input to the model. This strategic augmentation has proven effective in stabilizing FSL's performance, making it a more reliable method compared to traditional fine-tuning, which often struggles without extensive and carefully curated datasets. Through these enhancements, FSL demonstrates not only adaptability but also improved consistency in handling data diversity, thereby solidifying its position as a versatile tool in machine learning projects.

\textbf{RQ2: What is the cost of FSL vs. fine-tuning?}

The financial and resource-related implications of utilizing fine-tuning versus FSL are critically examined in our study. While fine-tuning has its merits, it often necessitates substantial computational resources and energy, leading to escalated costs, particularly when deploying large models on expansive datasets. Our findings indicate a significant reduction in these costs when adopting FSL, which minimizes reliance on voluminous data and extensive computational power. The cost-efficiency of FSL, combined with its competitive performance, firmly establishes it as a viable and practical alternative in settings constrained by resources.

The results collectively underscore the adaptability of FSL in handling the variability and complexity of software testing environments where data scarcity is prevalent. This adaptability and lower operational costs make FSL an attractive approach for integrating machine learning into the software development lifecycle, particularly in settings where rapid deployment and cost efficiency are priorities. In contrast, fine-tuning, although potentially more accurate in ideal conditions with abundant and diverse data, often requires extensive preprocessing and data curation efforts. These prerequisites can significantly delay deployment and escalate costs, making fine-tuning less suitable for environments that demand quick turnaround and cost-effective solutions. Consequently, while fine-tuning remains a powerful tool for scenarios well-endowed with resources, FSL's efficiency and flexibility offer critical advantages in more constrained settings, promoting its adoption as a practical alternative in the evolving landscape of software development.

However, in scenarios where an organization may have many projects that require flaky test detection and classification there will be a point where fine-tuning becomes a more cost effective option. Specifically, when the sum of the individual cost of training FSL for each project exceeds the cost of fine-tuning a generalizable model.

\section{Threats to Validity}

Several steps were taken to mitigate threats to validity in our research:

\begin{itemize}
    \item We utilized the third-party publicly available IDoFT and FlakyCat datasets to maintain transparency and reproducibility as well as avoid potential bias. However, these datasets may not capture the full diversity of flaky test scenarios encountered in all software projects in particular closed source projects which are not represented in either data set.
    \item We applied the same data pre-processing and evaluation metrics across all models in an effort to achieve a fair comparison between fine-tuning and few-shot learning approaches. Recognizing the class imbalance, where flaky tests are more prevalent, our use of augmentation and the use of weighted average helped prevent biased predictions towards the majority class. 

\item We based our hyperparameter settings on standard configurations, and although further tuning could potentially enhance model performance, we believe our comparative conclusions between FSL and fine-tuning in the context of flaky test detection and classification remain robust. 
\end{itemize}
By standardizing our evaluation procedures, and ensuring accessibility of our computational resources, we have aimed to minimize threats and strengthen the validity of our study. Despite best efforts several threats need to be acknowledged. First, the variability of model performance across different projects, as evidenced by lower F1-scores in certain categories like the Adyen project, suggests that some types of flakiness may require more specialized handling and it is unclear if the flakiness distribution across the data sets used is consistent with flaky tests in general. Second, while data augmentation techniques like SMOTE and code mutation enhance dataset diversity, they may introduce subtle biases. These biases, discussed in our recent work~\cite{riddhi2025}, highlight the need for careful augmentation design and further study. Third, our observations and conclusions regarding the performance and cost of FSL and fine-tuning approaches to the detection and classification of flaky tests do not necessarily reflect broader trends in these approaches when applied to other software development tasks. More experimentation is needed to generalize these observations with confidence.

\section{Conclusion \& Future Work}
\label{sec:Conclusion}

In this study, we observe that FSL, as implemented in the FlakyXbert model, is particularly effective in environments with sparse data, leveraging fewer examples to generalize well across flakiness categories. Fine-tuning, requiring more extensive data, excels in scenarios with diverse data by adapting to a broader range of features but demands greater computational resources and training time. The choice between FSL and fine-tuning hinges on balancing trade-offs between data availability, computational efficiency, and adaptability to diverse flaky test characteristics.

Beyond flaky testing, our analysis highlights the importance of contextualized assessments of LLM research in Software Engineering, as methods like fine-tuning and FSL can excel in complementary ways. In the future, we aim to adapt FlakyXbert for other classification tasks involving tests and source code, such as mutation test classification.

Further investigations should examine the scalability of FSL across diverse datasets and extend its applications to software engineering problems like bug detection, code optimization, and compliance checking. Future research could also explore hybrid approaches that combine fine-tuning for broad generalization with FSL for task-specific adaptability.

\balance
 \bibliographystyle{IEEEtran}

\bibliography{references}
\end{document}